# Thermoelectric terahertz photodetectors based on selenium-doped black phosphorus flakes

Leonardo Viti,*[a] Antonio Politano [b], Kai Zhang [c], Miriam Serena Vitiello [a]

Chemical doping of bulk black phosphorus is a well-recognized way to reduce surface oxidation and degradation. Here, we report on the fabrication of terahertz frequency detectors consisting of an antenna-coupled field-effect transistor (FET) with an active channel of Se-doped black phosphorus. Our devices show a maximum room-temperature hole mobility of 1780 cm$^2$V$^{-1}$s$^{-1}$ in a SiO$_2$-encapsulated FET. A room-temperature responsivity of 3 V/W was observed, with a noise-equivalent power of 7 nWHz$^{-1/2}$ at 3.4 THz, comparable with the state-of-the-art room-temperature photodetectors operating at the same frequency. Therefore, the inclusion of Se dopants in the growth process of black phosphorus crystals enables the optimization of the transport and optical performances of FETs in the far-infrared with a high potential for the development of BP-based electro-optical devices. We also demonstrate that the flake thickness can be tuned according to the target application. Specifically, thicker flakes (>80 nm) are suitable for applications in which high mobility and high speed are essential, thinner flakes (<10 nm) are more appropriate for applications requiring high on/off ratios, while THz photodetection is optimal with flakes 30-40 nm thick, due to the larger carrier density tunability.

## Introduction

Terahertz (THz) waves fall in the frequency spectrum between 0.1 THz and 10 THz, a region that has long been referred to as the THz *gap*, due to the demanding development of mature technologies for the generation, detection and manipulation of THz frequency beams. In the last decade, a flourishing of active and passive optoelectronic devices have appeared along with a number of disruptive applications spanning from medicine to biology, security, spectroscopy, astronomy and real time tomography [1,2]. In this latter context, the development of suitable imaging systems for in-situ and ex-situ applications has seen a recent boost, thanks to the broad development of room-temperature (RT), fast and scalable THz complementary metal-oxide semiconductor (CMOS) detectors, which foreshadows the production of a new-generation of cost-effective THz cameras [3-5]. Field-effect transistors (FETs), Schottky diodes and bolometers can be integrated in CMOS focal plane arrays (FPAs), ensuring attractive performances in the single-pixel configuration, i.e. noise equivalent powers (NEPs) < 5 nWHz$^{-1/2}$ and response times of 20 ns [1]. Amongst these architectures, FETs provide some clear advantages for THz photodetection, namely the inherent scalability and the combination of a fast response and high frequency operation (up to 22 THz) [6], very differently from Schottky diodes, whose performances are strongly affected by parasitic capacitances.

Beside CMOS technology, a great interest has recently emerged for nano-systems exploiting low-dimensional materials, which, while offering new twists for the understanding of novel quantum phenomena, can also ensure more practical technological advantages, such as the scalability and the compatibility with silicon for cost-effectiveness. In this context, single pixel FETs operating as RT THz photodetectors have been demonstrated, exploiting different two-dimensional (2D) layered materials like graphene [7-10], black phosphorus (BP) [11,12], van der Waals heterostructures [13] and topological insulators [14] or one-dimensional (1D) nanosystems such as carbon nanotubes [15] and semiconductor nanowires [16]. The rich physics involved in 2D materials can be exploited in a FET to engineer the detection dynamics from scratch. By playing with the device architecture and with the material geometry one can indeed selectively trigger the bolometric effect, the photo-thermoelectric effect (PTE) and/or the overdamped plasma-wave effect (OPW). The major difference between the three mechanisms is related to geometrical symmetry. The bolometric effect takes place when a modulation of the channel conductance occurs as a consequence of the homogeneous heating of the device. This effect can take place in any material whose conductivity $\sigma$ varies with temperature. If a voltage difference is applied through the material, the modulation of $\sigma$ results in a detectable modulation of current, *i.e.* in a generated photocurrent. Thus, the bolometric effect can occur in a symmetric FET, if a source (S) - to - drain (D) voltage ($V_{SD}$) is applied. On the other hand, both the PTE and OPW mechanisms need a certain degree of asymmetry in the detector structure. Indeed, the PTE is driven by a temperature gradient along the FET channel, and plasma-waves can be rectified inside the transistor when the THz field is coupled between S and gate (G) electrodes. The required asymmetry can be obtained either by asymmetrically coupling the free-space radiation to the active FET element by means of a planar dipole-like antenna [11], or by creating a *pn*-junction along the FET channel using adjacent gates [10], or by realizing asymmetrical boundary conditions at the FET extremities using doping or dissimilar contacting metals [8].

Owing to their peculiar electro-optical properties, graphene and BP are interesting platforms for the study and the activation of detection of THz radiation. In addition to the mentioned physical mechanisms, graphene can also behave like a ballistic rectifier, thanks to its efficient ballistic transport [17], reaching an NEP of 34 pWHz$^{-1/2}$ when the transmission coefficients in a four-ports device are modulated by THz frequency light [18]. Furthermore, BP can sustain photoconductive detection of free-

[a.] NEST, CNR-Istituto Nanoscienze and Scuola normale Superiore, Piazza San Silvestro 12, Pisa 56127, Italy.
[b.] Department of Physical and Chemical Sciences, University of L'Aquila, via Vetoio, 67100 L'Aquila, Italy.
[c.] i-Lab, Suzhou Institute of Nano-Tech and Nano-Bionics, Chinese Academy of Science, Suzhou 215123, China.
* Corresponding Author email: leonardo.viti@nano.cnr.it

space propagating 0.2 THz beams, when employing photo-gating effects in the near-infrared [12].

If on one hand the interest in graphene for THz applications mainly stems from its high mobility and gapless nature, on the other hand, the interest in BP is mainly triggered by its inherent in-plane anisotropy. The $sp^3$ hybridization of P atoms arranged in a 2D honeycomb lattice, leads to the formation of a puckered structure as a consequence of electrostatic repulsion [19]. The puckering is periodic along the armchair direction ([001], *c*-axis) and orthogonal to the zigzag direction ([100], *a*-axis). The in-plane crystallographic anisotropy affects the band structure, in turn inducing the anisotropy of its optical (linear dichroism [20]), electrical (mobility, effective mass [21]) and thermal (thermal conductance [22]) properties, opening the possibility to selectively activate different THz detection dynamics in a FET, by choosing the relative orientation between the armchair axis and the THz antenna axis (polarization axis) [23].

Another feature that makes BP extremely interesting among 2D layered materials is its dynamically tunable and direct bandgap via flake thickness control [24]. This effect is evident for ultrathin flakes (< 6 nm, 1 to 10 layers), whereas the bandgap saturates to about 0.3 eV above 20 layers (thickness > 11 nm).

However, despite these clear advantages over graphene, BP suffers from the lack of mechanical robustness and from a huge structural instability in ambient conditions, therefore requiring a controlled and demanding under-vacuum encapsulation immediately after (or during) its mechanical exfoliation. This is a clear disadvantage for a practical implementation of BP-based detectors in THz sensing systems.

In order to reduce surface oxidation and degradation of BP [25], a viable approach is represented by doping [26,27]. In particular, Te- and Se-doped BP flakes have enhanced transport performances and higher stability under ambient exposure with respect to pristine BP [27,28], offering a concrete perspective for a practical implementation of stable and robust BP-based devices. Along with the associated technological improvement, Se-doped few-layer BP flakes have recently been employed for the realization of active and passive optical components in the near infrared and in the visible frequency ranges [29,30], showing improved nonlinear optical properties and, more importantly, a significantly lower band gap (1.9 eV vs. 2.15 eV [30]) with respect to the undoped BP.

In order to take advantage of the aforementioned peculiarities, we devised antenna-coupled FETs where Se-doped BP flakes of different thicknesses were integrated as active channels and we studied their RT properties. First, we characterized the thickness and crystalline quality of the exfoliated flakes. Then we performed RT transport measurements in order to retrieve information about mobility ($\mu$) and charge density (*n*). Finally, we measured the THz photodetection performances of the devised detectors.

## Results and discussion

Se-doped BP flakes of different thicknesses, ranging from 8 nm to 80 nm, were transferred on Si/SiO$_2$ substrates via scotch-tape exfoliation in ambient conditions. The crystalline quality of the exfoliated flakes was assessed by polarized Raman spectroscopy (PRS). The Raman spectra (Fig, 1a), show the three main optically active BP vibration modes, $A_g^1$ (360 cm$^{-1}$), $B_{2g}$ (436 cm$^{-1}$) and $A_g^2$ (464 cm$^{-1}$), whose intensity ratios identify the crystallographic orientation of the BP flake, and the P-Se peak located at 187 cm$^{-1}$. The amplitude of this peak is expected to increase with the Se-content in the BP flakes [29].

Angle-resolved Raman spectroscopy was then used to determine the crystalline in-plane orientation of the flakes with respect to the substrate. To this end, we employed a linearly polarized incident radiation, with an excitation wavelength of 532 nm. It is known that an unambiguous definition of the flake orientation and a quantitative analysis of the Raman tensor can be obtained using polarized incident and scattered radiation [31-33], that is angle-resolved polarized Raman spectroscopy (ARPRS). However, even though in our angle-resolved experiment we have no control on the polarization of the scattered radiation, we observe that the $B_{2g}$ peak exhibits a periodicity of 90° (Figure 1b), with maxima occurring when the incident light is polarized at 45° between the armchair and zigzag axes. Thus, the 45° direction can be univocally defined even using a single polarizer. A typical polar plot of the amplitudes of the three main modes as a function of polarization angle is presented in Figure 1b.

Moreover, by taking into account the linear dichroism of BP and interference, we corrected the raw spectra by the enhancement factors described in Ref. [31] to assess the full orientation of the flakes. Note that the elements of the Raman tensor [33] cannot be unambiguously determined owing to the unknown polarization of the scattered light.

In order to make a consistent comparison between the devices realized for the present study, we adopted the following strategies: (*i*) BP flakes were exfoliated simultaneously on the same chip; (*ii*) we simultaneously fabricate all the devices on chip, employing parallel processing steps; (*iii*) the FETs architecture and the related integrated asymmetric THz antennas share identical geometry for all devices; (*iv*) in all cases the FET channel and the antennas were defined along the armchair crystalline direction.

Points (*i*) and (*ii*) ensure that the flakes suffer from the same exposure to the open environment and that the nanofabrication process is identical. It is worth mentioning that the reproducibility of the fabrication process in BP transistors is very demanding, since every lithographic step can impact the flake integrity, its crystalline quality and its surface cleanliness.

The geometry of the devised top-gated FETs is schematically represented in Figure 1c. The S and D FET electrodes were lithographically defined along the armchair BP direction at a relative distance of 1.0 μm, which defines the channel length (Lc). The size of the S and D ohmic contacts defines the channel width Wc = 1.5 μm. An 80 nm thick SiO$_2$ dielectric layer acting as top-gate oxide and encapsulating layer was then deposited via sputtering. The G electrode was lithographically patterned on the top surface at the center of the FET channel, over a length Lg = 0.5 μm = Lc/2.

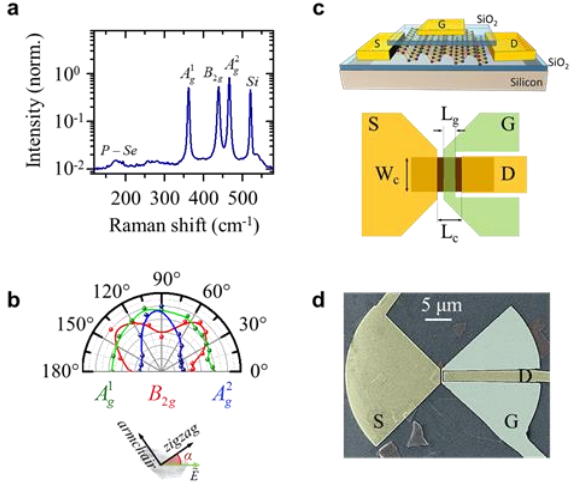

**Figure 1. Device Fabrication.** a) Raman spectrum collected along the armchair direction measured with a 532 nm excitation wavelength (logarithmic scale). Along with the three main BP modes, the Raman peaks of silicon (520 cm$^{-1}$) and of the *P-Se* phase (187 cm$^{-1}$) are visible. b) Amplitude of the three main Raman modes (A$_g^1$, green, B$_{2g}$, red, A$_g^2$, blue) plotted as a function of the angle (α) between the laser polarization and the zigzag direction. The amplitude of the B$_{2g}$ mode clearly shows maxima along the directions at 45° with respect to the main axes. c) Schematic diagram of the top gated BP-FET. The main geometrical dimensions are: W$_c$ = 1.5 µm, L$_c$ = 1.0 µm, and L$_g$ = 0.5 µm. A SiO$_2$ layer of 80 nm was used as gate dielectric. d) False color SEM image of a Se doped BP THz detector.

We here employed a split bow tie asymmetric THz antenna, as shown in the false colour scanning electron microscope (SEM) image of Figure 1d. The S and G electrodes were shaped as the two opposite halves of the bow tie, with a flare angle of 90° and a bow radius of 20 µm. At 3.4 THz and for normal incidence, this geometry gives a field enhancement of ~ 40 between the S and G branches of the antenna, when the electromagnetic wave, polarized parallel to the antenna axis, reaches the detector from the air side. This is estimated via a finite element method (FEM) simulation of the antenna architecture through a commercial software (Comsol Multiphysics).

Owing to the asymmetrical antenna geometry, THz detection is expected to be mediated either by the PTE effect or by the OPW mechanism, or through the simultaneous action of both mechanisms [23]. However, owing to the low acoustic photon speed along the armchair direction, that we selected on purpose, a dominant action of the PTE mechanism can be envisaged [34].

We fabricated five devices on the same chip. In the following, we will refer to samples *A*, *B*, *C*, *D* and *E*. The five samples only differ for the thickness of the BP flake that was measured by atomic force microscopy (AFM) before the fabrication of FETs. From the two dimensional AFM maps (top panels, Figs 2a-e) we extrapolated: 8 nm (*A*), 23 nm (*B*), 35 nm (*C*), 65 nm (*D*) and 80 nm (*E*), respectively.

The chip was then mounted on a dual inline and the samples were electrically characterized in air at RT using two *dc* voltage generators to apply source-to-drain ($V_{SD}$) and source-to-gate ($V_G$) voltages. The measured resistance curves ($R_c$) as a function of $V_G$ are plotted in Figure 2 (bottom panels).

All devices exhibit a *p*-doping at zero gate voltage. However, compared to previously reported experiments on undoped BP FETs, fabricated following an identical procedure [23], we found that Se-doped samples typically have a lower carrier density at $V_G$ = 0 V, i.e. they are less doped. Indeed, the estimated density of holes is of the order of $10^{17}$ cm$^{-3}$ for all tested samples, whereas it was ~$10^{19}$ cm$^{-3}$ in the undoped BP FETs [23]. This effect can be understood by the electron concentration introduced by Se-doping. Selenium acts as a donor for BP, thus providing extra electrons to a typically *p*-type semiconductor, eventually decreasing the concentration of majority carriers (*i.e.* holes). A progressive increase of the selenium content during the growth process is expected to bring the charge neutrality point (CNP) closer to $V_G$ = 0 V. Moreover, the inclusion of selenium has been proven to slightly lower the bandgap of BP [29]. As a result, it is easier to obtain ambipolar transfer behavior for Se-doped BP, as we clearly unveiled in samples *C*, *D* and *E* where the transition from hole-current to electron-current can be clearly seen (Figs. 2c-e).

The field-effect mobility ($\mu_{FE}$) of the five samples was evaluated from the transfer characteristics curves of the FETs via the equation:

$$\mu_{FE} = \frac{g_m L_g^2}{C_{gc} V_{SD}}, \qquad (1)$$

where $g_m$ is the maximum FET transconductance obtained from a linear fit to the curves in Figure 2 and $C_{gc}$ = 400 aF is the geometrical gate-to-channel capacitance, extracted via tri-dimensional FEM simulations and which is assumed to be identical for all the samples. As reported in ref. [11], the total capacitance of a top-gated BP-FET with comparable size can be safely approximated with the geometrical capacitance $C_{gc}$, neglecting the contributions given by the trap and quantum capacitances.

The $\mu_{FE}$ values retrieved for samples *A-E* are summarized in Table 1. The maximum value for hole mobility ($\mu_h$) is obtained for sample *E*, $\mu_h$ = 1780 cm$^2$V$^{-1}$s$^{-1}$, showing an increase of more than a factor 3 with respect to previously reported SiO$_2$-BP-SiO$_2$ heterostructures measured at RT [11] and a factor of ~ 2 larger than that extrapolated in BP structures exploiting oxide/polymeric encapsulation [35]. This can be attributed to the combined effect of low carrier density and smaller bandgap. Another important parameter of a FET is the ratio between the on-state current ($I_{ON}$) and the off-state current ($I_{OFF}$), which is directly linked to the carrier density tuning efficiency provided by the G electrode.

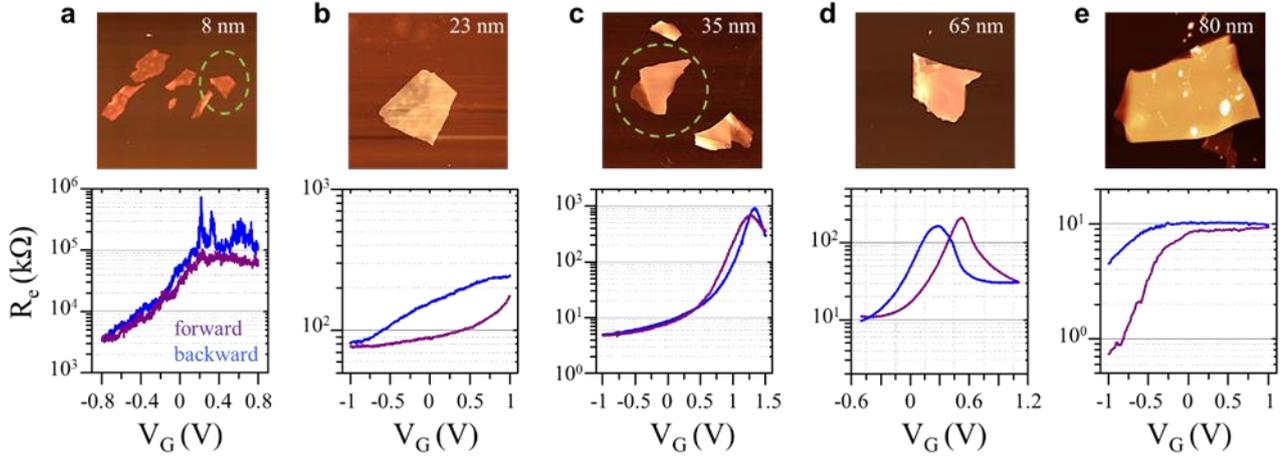

**Figure 2. Electrical Characterization.** a-e) Top panels: AFM images of the Se-doped BP flake measured for sample *A* (a), *B* (b), *C* (c), *D* (d) and *E* (e). Bottom panels: room temperature resistance as a function of applied top-gate voltage, swept in the forward and backward directions for samples *A* (a), *B* (b), *C* (c), *D* (d) and *E* (e). An ambipolar behaviour is clearly visible for samples *C* and *D*.

The obtained values of $I_{ON}/I_{OFF}$ are summarized in Table 1. As reported in ref. [24], in undoped BP-based ambipolar FETs, $I_{ON}/I_{OFF}$ should be larger for thinner flakes. Here we observe that this trend is fulfilled for the devices that show ambipolar transport in the tested gate voltage range: samples *C*, *D* and *E*. Devices *A* and *B* do not follow this rule, because of the limited gate sweep that was used in our experiments (motivated by the presence of a thin top-gate oxide layer employed to maximize the gate-to-channel capacitance per unit area). Notably, sample *C* exhibits a two orders of magnitude carrier density variation in a very small gate voltage sweep: when $V_G$ is swept from 1.25 V to 0.25 V, *n* changes from 2.5 x $10^{15}$ cm$^{-3}$ to 2.5 x $10^{17}$ cm$^{-3}$.

In order to characterize the RT THz response of the five samples, we employed a quantum cascade laser (QCL) operating in pulsed mode at a wavelength $\lambda_{qcl}$ = 88 μm (3.4 THz). A detailed description of the experimental set-up is given in the *Experimental* section. The THz beam was focused in a circular spot of radius 100 μm by two lenses (f/2). The average power at the detector position was $P_t$ = 80 μW. We used a lock-in amplifier, along with a low-noise pre-amplification stage (gain *G* = 1000), to measure the photovoltage (*Δu*) between D and S. The mathematical relation between *Δu* and the signal read by the lock-in (*LIA*) is given by: *Δu* = 2.2\**LIA*/*G* [34]. The voltage responsivity ($R_v$) was then determined by normalizing the photovoltage with respect to the optical power impinging on the detector, that is:

$$R_v = \frac{\Delta u}{P_t} \cdot \frac{A_{spot}}{A_{diff}} \quad (2)$$

Here the beam spot area ($A_{spot}$) is calculated as the circular area of the intensity distribution, with radius given by its half width at half maximum (HWHM). $A_{diff}$ represents the diffraction-limited area that we assumed to be equal to $\lambda^2/4$. Figures 3 (a-e) show the plots of $R_v$ recorded at RT as a function of $V_G$ for the five samples. In all cases, a noticeable photodetection signal has been measured. Two of the samples outperform the others in terms of figure of merits: $|R_v|$(max) = 2.9 V/W for sample *C* and $|R_v|$(max) = 1.5 V/W for sample *D*. Interestingly, these are the two samples with the ideal combination of high mobility and high carrier density tunability (high $I_{ON}/I_{OFF}$ ratio). The experimental results provide clear evidence of the physical mechanism underlying the detection. We will restrict our discussion to samples *C* and *D*, since they show a clear dependence of the responsivity $R_v$ from $V_G$. As discussed in the *Introduction*, when the THz field is asymmetrically coupled between the S and G electrodes by a planar antenna, two mechanisms are expected to take place, *i.e.* the PTE and OPW effects. In the latter case, by sweeping the gate voltage ($V_G$), the photoresponse is expected to switch its sign when $V_G$ corresponds to the transition of the FET from *n*-type to *p*-type (CNP).

**Table 1.** Summary of estimated electrical parameters for the five BP-FETs

| Sample | Thickness (nm) | $\mu_h$ (cm$^2$V$^{-1}$s$^{-1}$) | $\mu_e^1$ (cm$^2$V$^{-1}$s$^{-1}$) | $I_{ON}/I_{OFF}$ |
|---|---|---|---|---|
| A | 8 | 2 | - | 50 |
| B | 23 | 42 | - | 3 |
| C | 35 | 536 | 102 | 225 |
| D | 65 | 910 | 604 | 22 |
| E | 80 | 1780 | - | 13 |

[1] $\mu_e$ is the electron mobility, it is calculated only for samples showing ambipolar transport.

The latter transition can be clearly unveiled in the transport characteristics, since it appears as a negative-to-positive sign change in the derivative of the transfer curve; this in turn reflects into a single sign change in the photovoltage $\Delta u_{opw} \sim 1/\sigma \cdot \partial \sigma / \partial V_G$ [34]. The $R_v$ curves reported in Figure 4c,d show two evident sign changes (the sign change occurring in Figure 4d for $V_G$ = -0.7 V is due to a leakage in the gate electrode), in clear disagreement with the prediction of the overdamped plasma-wave effect.

On the other hand, the PTE mechanism could explain the double sign change in the detector responsivity. The PTE voltage ($\Delta u_{PTE}$) is the result of a temperature gradient $\Delta T$ along the FET channel. Hot carriers accumulate within the ungated region between S and G as a consequence of the THz-induced field provided by the antenna. Conversely, electrons remain colder in the region underneath the D and G electrodes. The photovoltage given by the Seebeck effect can be written as $\Delta u_{PTE} = (S_{bG} - S_{bU}) \cdot \Delta T$, where $S_{bG}$ and $S_{bU}$ are the Seebeck coefficients of the gated portion and ungated portions of the channel, respectively. The Seebeck coefficient $S_{bU}$ (dashed horizontal line in Figure 3f) is not affected by $V_G$ and can be ascribed to the slightly p-doped BP in the ungated regions. Vice versa $S_{bG}$ changes with $V_G$ following the Mott equation [11]:

$$S_{bG} = -\frac{\pi^2 k_B^2 T}{3e} \left( \frac{1}{\sigma} \cdot \frac{\partial \sigma}{\partial V_G} \right) \frac{\partial V_G}{\partial E_f} \quad (3)$$

Where $k_B$ is the Boltzmann constant, $e$ is the electron charge and $E_f$ is the Fermi energy. Figure 3f shows the calculated $S_{bG}(V_G)$ extracted from the resistance curve of sample C (Figure 2c), by using equation (3) and applying the same method described in ref. [11]. $S_{bG}$ is negative when the gated region is n-type and positive when the gated region is p-type. The PTE voltage dependence from $V_G$ follows the $S_{bG}$ trend and it is zero when $S_{bG} = S_{bU}$, that is when $V_G \sim 0$ V and when $V_G \sim$ CNP. In the range of gate voltages comprised between these two values $S_{bG} > S_{bU}$ and $\Delta u_{PTE}$ is positive, elsewhere $S_{bG} < S_{bU}$ and $\Delta u_{PTE}$ is negative, thus giving rise to the two sign changes observed in the responsivity curves (see Figures 3c and 3f for a visual comparison between experimental and predicted $R_v$). As expected [23], when the antenna axis is parallel to the armchair direction of the BP flake, in which the acoustic phonon speed (i.e. the thermal conductivity) is lower than along the zigzag axis, the PTE mechanism dominates THz detection. From the evaluation of the Seebeck coefficient of sample C, we can calculate the temperature increase along the FET channel as the ratio between $\Delta u_{PTE}$ and the difference $S_{bG} - S_{bU}$, when the responsivity is maximum, i.e. for n-type doping. We obtained $T_{hot} - T_{cold}$ = 0.5 K. To better quantify the detection performances of our devices, we determined the minimum NEP. This figure of merit is defined as the ratio between the noise spectral density (NSD, units V·Hz$^{-1/2}$) and the detector responsivity and represents the minimum incident power required to achieve a unitary signal-to-noise ratio within a bandwidth of 1 Hz. Since a modulation frequency of 1.333 kHz was used to trigger our lock-in detection scheme, the major contribution to NSD is given by the thermal fluctuations within the FET channel (the 1/f term can be neglected [27]), which can be calculated using the well-known Nyquist theorem: $\boldsymbol{V_{noise} = \sqrt{4k_B T R_c}}$. Then NEP$_{min}$ = $V_{noise}/R_v$ = 7 nW/√Hz for sample C, NEP$_{min}$ = 23 nW/√Hz for sample D and NEP$_{min}$ > 100 nW/√Hz for the other three samples. The minimum NEP obtained in sample C is comparable with the best values reported to date in the undoped BP photodetectors operating at the same frequency [34].

It is worth mentioning that, the reported data have to be considered as a first demonstration of a thermoelectric Se-doped BP photodetector. Indeed, in order to fully take advantage of the thermoelectric effect, the device geometry and the antenna shape can be largely improved to reduce heat dissipation to the environment, thus optimizing the generation of a longitudinal thermal gradient along the channel under THz illumination.

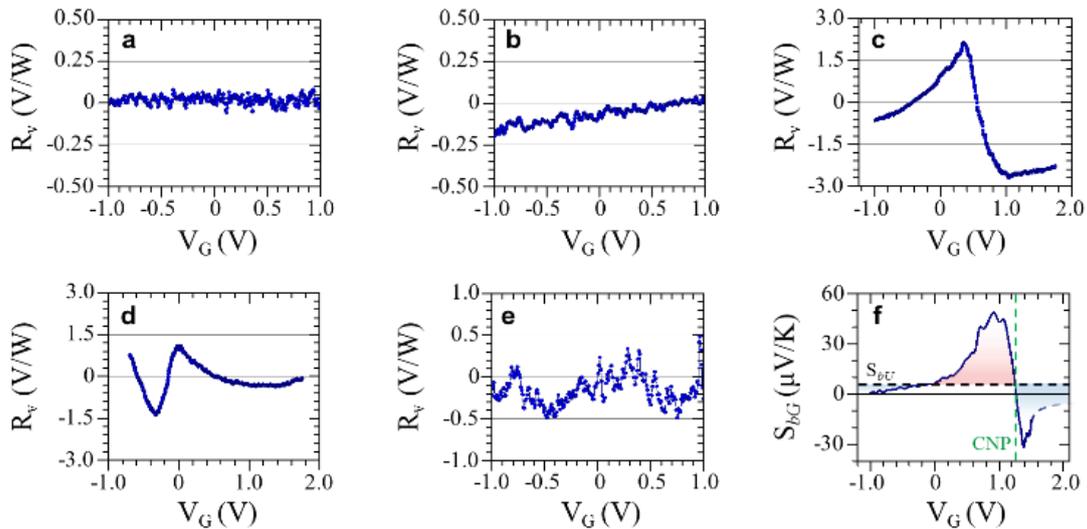

**Figure 3. Optical Characterization**. a-e) Responsivity plotted as a function of the gate voltage, measured at 3.4 THz, and at RT for device A (a), B (b), C (c), D (d) and E (e). The voltage signal was taken at the D electrode of the FETs, while keeping S grounded. The optical intensity impinging on the detectors was 0.25 Wcm$^{-2}$. f) Evaluation of the Seebeck coefficient of the gated area ($S_{bG}$) as a function of $V_G$, extrapolated from equation (3). The dashed horizontal line represents the value of the Seebeck coefficient of the ungated areas ($S_{bU}$). The dashed vertical line (green) indicates the charge neutrality point (CNP). The thermoelectric photovoltage is proportional to the difference $S_{bG} - S_{bU}$, hence it is negative for $V_G$ < 0 V (blue-shaded area), positive for 0 V < $V_G$ < CNP (red shaded area) and negative for $V_G$ > CNP, i.e. n-doping (blue shaded area).

The comparison between the electrical and optical characteristics of the five photodetectors shows that the responsivity, the mobility and the on/off current ratios in photodetectors characterized by different layer thicknesses follow a specific trend (Figure 4).

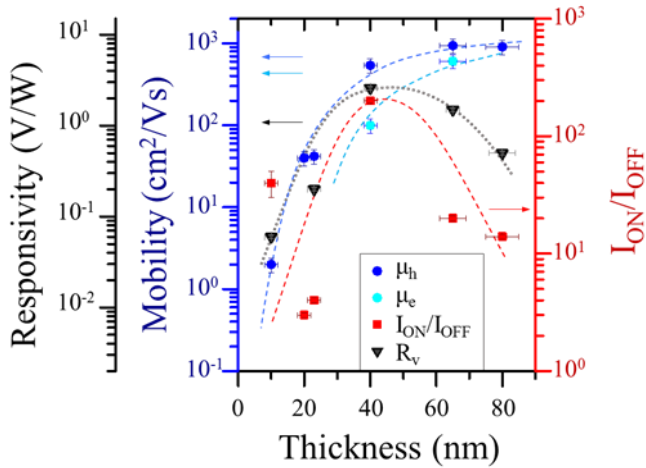

**Figure 4. Comparison of the electrical and optical performances.** (1st left vertical axis) Room-temperature responsivity $R_v$ (black triangles). (2nd left vertical axis) Mobility of holes (blue circles) and electrons (light blue circles). (right vertical axis) $I_{ON}/I_{OFF}$ ratio (red squares), as a function of the flake thickness. Dashed and dotted lines are guides for the eye.

Thicker flakes show higher mobility with respect to thinner ones, at the expenses, however, of a reduced $I_{ON}/I_{OFF}$ ratio: $\mu_h$ for sample *E* is 1780 cm$^2$/Vs, but the $I_{ON}/I_{OFF}$ ratio is only 13. On the other hand, we observe the complete closing of electrical transport only for sample *A*, because its thickness (8 nm) is lower than the out-of-plane screening length (~10 nm) [36], *i.e.* the thickness at which the interlayer electrostatic screening cancels the effect of the applied gate voltage.

The comparison of the optical performances shows that the responsivity is larger for those devices (*C* and *D*) that show the highest $I_{ON}/I_{OFF}$ ratio and also a good mobility. In a standard FET photodetector a high switching efficiency (represented by a high $I_{ON}/I_{OFF}$) is usually preferred to a highest mobility, which plays a secondary role in the thermoelectric response. Indeed, the curve for $R_v$ (black dotted line) in Figure 4 follows the trend of the $I_{ON}/I_{OFF}$ curve (red dashed line).

## Experimental

### Growth
Selenium-doped BP crystals with controllable doping contents were obtained by the mineralizer-assisted gas-phase transformation method described in Ref. [29]. The weight percentage (wt%) of selenium employed for the crystal used in this work is 1.6%.

### Fabrication
BP flakes were exfoliated on an intrinsic silicon substrate capped with thermally grown SiO$_2$ (300 nm). The scotch-tape exfoliation was performed in ambient condition, starting from as-grown crystals of lateral size 5 mm. Five different flakes were then selected and their thickness was characterized by atomic force microscopy (thickness resolution < 1 nm and lateral resolution ~ 20 nm). The chip was then covered with polymethyl methacrylate (PMMA) and stored in vacuum. The in-plane orientation of the flakes with respect to the host substrate was determined by polarized Raman spectroscopy, using a Renishaw (InVia) system, equipped with a frequency doubled Nd:YAG laser emitting a wavelength of 532 nm. S and D electrodes were then realized using aligned EBL and metal evaporation of 10/100 nm of Cr/Au. Immediately after metal lift-off, the chip was covered with an 80 nm thick SiO$_2$ layer obtained by Ar sputtering. The encapsulation prevents further oxidation and degradation of the flakes. We estimate that between the moments of exfoliation and encapsulation, the flakes are exposed to air for a total time of 1 hour (roughly corresponding to the AFM time). The top-gate contact was then realized with EBL and metal evaporation (Cr/Au, 10/100 nm).

### Terahertz optical setup
THz light was generated by a single-plasmon-waveguide QCL operating at 3.4 THz. The QCL was cooled down by a Stirling cryocooler with a heat sink temperature of 27 K, resulting in an approximate lattice temperature of 80 K. The cryostat was closed with a high-density polyethylene (HDPE) window, which transmits the 80% of the power when the QCL was aligned with its center. The diverging THz beam was collimated and then focused by means of two tsupurica (*Picarin*) lenses with focal lengths 30 mm (collimating) and 50 mm (focusing), obtaining a focal spot of ~ 100 μm radius (HWHM) and an average intensity of 0.25 W/cm$^2$. The QCL was electrically pumped by a voltage square waveform with repetition frequency 40 kHz and duty cycle 4%, resulting in a pulse duration of 1 μs. In order to perform lock-in measurements a 1.333 kHz square wave envelope was used as a trigger for the pulse trains. This allowed using the 1.333 kHz as the reference signal for the phase-locked detection, thus reducing the noise level of the measurement setup.

## Conclusions

The inclusion of selenium in the growth process of black phosphorus crystals opens new possibilities for optimizing the transport and terahertz optical performances of FETs exploiting encapsulated Se-doped BP flakes, we have shown an increase in room-temperature field-effect mobility of a factor ~3 with respect to our previous results on similarly fabricated devices exploiting undoped BP.

PTE-mediated detection at 3.4 THz has been reported for the five fabricated devices. Among these, the best working one (thickness 35 nm, sample C) showed a RT $R_v$ = 3 V·W$^{-1}$ and a minimum NEP = 7 nW·Hz$^{-1/2}$. The key signature of the PTE mechanism has been identified in a double sign switch of the photovoltage as a function of $V_G$.

Eventually, the comparison between the devised samples allowed us to infer a *rule-of-thumb* for the selection of the optimal flake thickness according to the target application. When high mobility is required (e.g. in high on-state current, high switching speed applications) thicker flakes (> 80 nm) should be preferred. For complete FET closing with high $I_{ON}/I_{OFF}$, thinner flakes should be used (< 10 nm, *i.e.* below the out-of-plane screening length). For photodetection of THz frequency light, 30-40 nm thick flakes are preferable, owing to their combined high mobility and good tunability of carrier density.

Our results demonstrate that Se-doped BP represents a valid candidate for high-speed electronics requiring a high on-state current thanks to the increased mobility. In this context, major improvements to the results reported in this manuscript could be represented by the replacement of the $SiO_2$ encapsulation with vacuum hBN encapsulation [37], which has been recently demonstrated to give electron mobility > 5000 $cm^2V^{-1}s^{-1}$ in FET based on undoped BP.

## Conflicts of interest

There are no conflicts to declare.

## Acknowledgements

We acknowledge financial support from the EC Project Graphene flagship core project II (785219), and the ERC consolidator grant SPRINT (681379). M.S.V. acknowledges partial support from the second half of the Balzan Prize 2016 in applied photonics delivered to Federico Capasso.